\documentclass[%
superscriptaddress,
longbibliography,
twocolumn,
amsmath,
amssymb,
aps,
prl,
floatfix,
10pt,
]{revtex4-2}

\usepackage{graphicx}
\usepackage{dcolumn}
\usepackage{bm}
\usepackage{siunitx}
\usepackage{xcolor}
\usepackage[colorlinks=true, linkcolor=blue, urlcolor=blue, citecolor=blue]{hyperref}

\begin{document}


\title{
Analytical Forces from the Bethe-Salpeter Equation for Large-Scale Excited-State Relaxation
} 

\author{Yu Jin}
\email{yjin@flatironinstitute.org}
\affiliation{Pritzker School of Molecular Engineering, University of Chicago, Chicago, Illinois 60637, USA}
\affiliation{Initiative for Computational Catalysis, Flatiron Institute, New York, New York 10010, USA}
\author{Victor Wen-zhe Yu}
\affiliation{Materials Science Division, Argonne National Laboratory, Lemont, Illinois 60439, USA}
\author{Marco Govoni}
\affiliation{Department of Physics, Computer Science, and Mathematics, University of Modena and Reggio Emilia, Modena, 41125, Italy}
\author{Giulia Galli}
\email{gagalli@uchicago.edu}
\affiliation{Pritzker School of Molecular Engineering, University of Chicago, Chicago, Illinois 60637, USA}
\affiliation{Department of Chemistry, University of Chicago, Chicago, Illinois 60637, USA}
\affiliation{Materials Science Division, Argonne National Laboratory, Lemont, Illinois 60439, USA}

\date{\today}
\begin{abstract}
We present an efficient plane-wave implementation of analytical nuclear forces for electronic excited states described by the Bethe-Salpeter equation (BSE). The formulation combines density-matrix perturbation theory with a Lagrangian approach, and avoids both explicit empty-state summations and the response calculations for each atomic displacement, required by conventional approaches based on density functional perturbation theory. Together with GPU acceleration, these advances make BSE forces calculations tractable for solid-state systems containing hundreds of atoms. We demonstrate the method on two point defects with distinct dielectric environments: the nitrogen-vacancy center in diamond, where BSE and time-dependent density functional theory (TDDFT) yield consistent excited-state relaxations, and the carbon-dimer defect in two-dimensional hexagonal boron nitride, where the screened electron-hole interaction included in the BSE stabilizes the localized defect excitation and corrects the relaxation pattern predicted by semilocal TDDFT. These results establish a scalable framework for BSE-level studies of excited-state relaxation and vibronic coupling in heterogeneous condensed systems.
\end{abstract}

\maketitle

{\textit{Introduction}}---Excited-state (ES) phenomena in materials are often governed by the coupling between electronic and vibrational degrees of freedom. This coupling determines, for example, the vibronic structure of optical spectra of point defects~\cite{huang1950theory,alkauskas2014first,alkauskas2016tutorial} and the formation and relaxation of self-trapped excitons in perovskites~\cite{li2019self,li2025luminescence}. Hence, a theoretical microscopic description of these processes requires not only accurate calculations of excitation energies, but also of gradients of ES potential energy surfaces, or ES forces~\cite{lewis2021modeling,sun2023toward,seo2024first,gali2023recent,ping2021computational,xiong2026finding}. Density functional theory (DFT) based approaches, including $\Delta$SCF and time-dependent DFT (TDDFT), have been widely used to study ES relaxation in materials~\cite{gali2009theory,wang2019atomistic,jin2022vibrationally,jin2023excited}. However, both approaches are not designed to capture spatially inhomogeneous screening effects, especially when used with semilocal or conventional hybrid functionals~\cite{botti2007time,sun2021pros}. In addition, $\Delta$SCF is limited to single-determinant descriptions of excited states and often suffers from convergence difficulties~\cite{xiong2025delta}.

Many-body perturbation theory, in particular the $GW$ approximation combined with the Bethe-Salpeter equation (BSE), has become a standard high-level framework for describing neutral excitations in materials. However, applications of $GW$-BSE have largely focused on vertical excitation energies and optical absorption spectra, because the calculation of ES forces is substantially more demanding than that of energies. Early analytical-gradient formulations by Ismail-Beigi and Louie~\cite{ismailbeigi2003excited,ismailbeigi2005self}, recently revisited by Del Grande and Strubbe~\cite{del2025revisiting}, were implemented in a plane-wave basis and used density-functional perturbation theory (DFPT) to compute the derivatives of occupied and virtual Kohn-Sham (KS) orbitals with respect to each nuclear degree of freedom. This procedure requires solving $3N_{\text{atom}}$ self-consistent Sternheimer equations, making the calculations of BSE forces prohibitively expensive for large systems. Further, the formulations of Ref.~\citenum{ismailbeigi2003excited} introduced approximations to the derivatives of quasiparticle (QP) energies and screened Coulomb interactions. Recently, Villalobos-Castro \textit{et al.} introduced a $Z$-vector formulation that replaces the $3N_{\text{atom}}$ separate Sternheimer calculations by a single response equation~\cite{villalobos2023lagrangian}; this approach was further generalized by T\"{o}lle \textit{et al.} to obtain fully analytical nuclear gradients without approximations to the QP energies or screened Coulomb interactions~\cite{tolle2025fully}. These $Z$-vector implementations, however, have so far been formulated only in Gaussian basis sets for molecular systems and retain steep computational scaling, typically $\mathcal{O}(N^6)$ to $\mathcal{O}(N^7)$, hampering their applicability to large condensed-phase systems.

Excited-state (ES) structural relaxations in materials have also been estimated without the explicit calculation of ES forces, for example, by combining harmonic descriptions of the potential energy surface with shift-mode approximations~\cite{dai2024excitonic,dai2024theory,dai2025polarons,bai2024ab,yang2022novel}. These approaches are computationally efficient as they avoid explicit geometry relaxations in large supercells, but their accuracy is limited in cases where anharmonicity and high-order electron-phonon coupling play an important role, as well as by assumptions on the dominant relaxation coordinate. An accurate and scalable approach to compute BSE forces is therefore needed for predictive studies of ES relaxation in realistic solids.

In this Letter, we present an efficient plane-wave implementation of analytical BSE forces applicable to solids with hundreds of atoms, and we assess the performance of TDDFT and BSE based approaches for homogeneous and inhomogeneous dielectric environments.   Building on our previous work on TDDFT analytical forces~\cite{jin2022vibrationally,jin2023excited}, we use a $Z$-vector Lagrangian formulation to avoid separate orbital-response calculations for each nuclear degree of freedom. Our implementation in the open-source \textsc{WEST} code~\cite{govoni2015large,yu2022gpu,yu2026west} employs a low-rank representation of the dielectric matrix and avoids explicit evaluation of empty states. We show that while for the nitrogen-vacancy center (NV$^-$) in diamond, BSE and hybrid-functional TDDFT yield similar results on ES relaxations, an explicit treatment of the screened Coulomb interactions is necessary for a system with an inhomogeneous dielectric environment, such as the carbon dimer substitute (C$_\text{B}$C$_\text{N}^0$) in two-dimensional hexagonal boron nitride (2D-hBN).

{\textit{Theoretical framework}}---Within density-matrix perturbation theory (DMPT), the BSE in the Tamm-Dancoff approximation (TDA) for a spin-restricted system can be written as~\cite{rocca2010ab,roca2012solution,nguyen2019finite,jin2023excited,yu2024gpu}
\begin{equation}
    \left( \mathcal{D} + 2 \mathcal{K}^{\text{1e}} - \mathcal{K}^{\text{1d}} \right) \mathcal{A}_s = \omega_s \mathcal{A}_s ,
\label{eq:bse}
\end{equation}
where $\omega_s$ is the vertical excitation energy of the $s$-th excited state. The eigenvector $\mathcal{A}_s \equiv \{|a_{s,v}\rangle, v=1,\ldots,N_{\mathrm{occ}}\}$ represents response orbitals associated with the occupied KS orbitals $\{|\varphi_v\rangle\}$. In the following, we suppress the state index $s$ for simplicity. The action of the three operators in Eq.~\eqref{eq:bse} is
\begin{subequations}
\begin{align}
    \left(\mathcal{DA}\right)_v &= \hat{P}_{c} \big( \hat{H}^{\text{QP}} - \varepsilon_{v}^{\text{QP}} \big) | a_{v} \rangle, \label{eq:d} \\
    \left(\mathcal{K}^{\text{1e}} \mathcal{A} \right)_v &= \hat{P}_c  \int d \mathbf{r}' v_c(\mathbf{r},\mathbf{r}') \Delta \rho (\mathbf{r}') | \varphi_v \rangle, \label{eq:k1e} \\
    \left(\mathcal{K}^{\text{1d}} \mathcal{A} \right)_v &= \hat{P}_c \sum_{v'} \int d \mathbf{r}' W(\mathbf{r},\mathbf{r}') \varphi_{v'}^{\ast} (\mathbf{r}') \varphi_{v} (\mathbf{r}') | a_{v'} \rangle.
\label{eq:k1d}
\end{align}
\end{subequations}
Here, $\hat{H}^{\mathrm{QP}}$ is the QP Hamiltonian, $\varepsilon_v^{\mathrm{QP}}$ is the QP energy associated with $|\varphi_v\rangle$, $v_c$ is the bare Coulomb potential, and $\hat{P}_c$ projects onto the unoccupied subspace. The transition density is $\Delta\rho(\mathbf{r})=\sum_v \varphi_v^{\ast}(\mathbf{r})a_v(\mathbf{r})$. The statically screened Coulomb interaction $W(\mathbf{r},\mathbf{r}')$ is evaluated using the projective dielectric eigenpotential (PDEP) technique within the
random phase approximation~\cite{wilson2008efficient,wilson2009iterative,nguyen2012improving,pham2013gw},
\begin{equation}
W(\mathbf{r},\mathbf{r}') = v_c(\mathbf{r},\mathbf{r}') + \frac{1}{\Omega} \sum_{a=1}^{N_{\mathrm{pdep}}} \widetilde{\phi}_a(\mathbf{r}) \frac{\lambda_a}{1 - \lambda_a} \widetilde{\phi}^{\ast}_a (\mathbf{r}'),
\end{equation}
where $\Omega$ is the volume of the simulation cell, $\widetilde{\phi}_a(\mathbf{r})=\int d\mathbf{r}' v_c^{1/2}(\mathbf{r},\mathbf{r}')\phi_a(\mathbf{r}')$, and $\lambda_a$ and $|\phi_a\rangle$ are the leading eigenvalues and eigenvectors of the symmetrized irreducible density-density response function. This formulation of the BSE avoids explicit construction and inversion of the dielectric matrix, and removes the need to converge the excitation energy with respect to a large number of occupied and unoccupied orbitals compared to conventional electron-hole-basis formulations~\cite{rohlfing2000electron}.

The ES forces are the sum of the ground-state (GS) forces and the negative of the derivatives of $\omega$ with respect to nuclear coordinates; the latter can be evaluated using an extended Lagrangian formalism~\cite{hutter2003excited,zhang2015subspace} with the resulting expression written as
\begin{equation}
    \frac{d \omega}{d R} = \int d \mathbf{r} \frac{\partial V_{\text{ext}}(\mathbf{r})}{\partial R} \big[\Delta \rho^{(x)}(\mathbf{r}) + \Delta \rho^{(z)}(\mathbf{r}) \big].
\end{equation}
Here, $\partial V_{\text{ext}}(\mathbf{r}) / \partial R$ denotes the derivative of the external potential, represented by the pseudopotential derivative in our implementation. The unrelaxed differential density of the excited state is
\begin{equation}
    \Delta \rho^{(x)}(\mathbf{r}) = \sum_v a_v^{\ast}(\mathbf{r})a_v(\mathbf{r}) - \sum_{vv'} \varphi_v^{\ast}(\mathbf{r})\varphi_{v'}(\mathbf{r})
\langle a_{v'} | a_v \rangle ,
\end{equation}
and the contribution from KS orbital relaxation is
\begin{equation}
    \Delta \rho^{(z)}(\mathbf{r}) = \sum_v \left[ Z_v^{\ast}(\mathbf{r})\varphi_v(\mathbf{r}) + \varphi_v^{\ast}(\mathbf{r})Z_v(\mathbf{r}) \right].
\end{equation}
The response orbitals $Z_v(\mathbf{r})$ are obtained by solving the Handy--Schaefer $Z$-vector equation~\cite{handy1984evaluation},
\begin{equation}
\label{eq:z-vector}
    \left( \mathcal{D}^{\mathrm{KS}} + 2 \mathcal{K}^{\mathrm{1e},\mathrm{KS}} - \mathcal{K}^{\mathrm{1d},\mathrm{KS}} + 2 \mathcal{K}^{\mathrm{2e},\mathrm{KS}} - \mathcal{K}^{\mathrm{2d},\mathrm{KS}} \right) \mathcal{Z} = -\mathcal{U},
\end{equation}
where $\mathcal{U}$ denotes the set of quantities obtained from the derivative of $\omega$ with respect to the occupied KS orbitals. The explicit form of the extended Lagrangian, $\mathcal{U}$, and the KS response kernels in Eq.~\eqref{eq:z-vector} are given in Sec.~SI of the Supplemental Material (SM~\cite{SM}). Compared with DFPT-based formulations, which require solving $3N_{\mathrm{atom}}$ Sternheimer equations self-consistently to obtain the BSE forces, the $Z$-vector formulation requires only the solution of one Sternheimer-like response equation. We emphasize that this reduction in complexity is particularly important when using DFT with hybrid functionals to compute initial single-particle orbitals, for which DFPT derivatives are substantially more expensive and less explored, in plane-wave implementations, than the respective computations with semi-local functionals.

The formalism above provides the basis for analytical BSE forces. In the present implementation, the evaluation of $\mathcal{U}$ uses two approximations: (i) the derivative of the QP Hamiltonian is approximated by that of the KS Hamiltonian, $\delta \hat{H}^{\mathrm{QP}}/\delta \langle \varphi_v| \approx \delta \hat{H}^{\mathrm{KS}}/\delta \langle \varphi_v|$,
which is equivalent to using a fixed scissor operator to correct the QP energies~\cite{villalobos2023lagrangian,alrahamneh2025excited}; (ii) the derivative of the screened Coulomb interaction $\delta W/\delta \langle \varphi_v|$ is set to zero. We validate these approximations for the lowest singlet excited state of carbon monoxide (see End Matter for details). 

We further verify our implementation by comparing analytical and finite-difference forces at the same level of approximation, finding excellent agreement for both semi-local and hybrid-functional DFT starting points (see Sec.~SII~\cite{SM}), and we benchmark optimized ES geometries for several small molecules. The resulting geometries closely agree with recent fully analytical BSE forces calculations~\cite{tolle2025fully} and, in some cases, are closer to high-level wave-function benchmarks, possibly due to differences in the mean-field starting points used in the two calculations (see Sec.~SIII~\cite{SM}).

The computational cost of our implementation is dominated by the construction of the projective dielectric eigenpotential (PDEP) basis, which scales as $\mathcal{O}(N^4)$~\cite{govoni2015large}. To make BSE forces calculations practical for large supercells, we combine several algorithmic accelerations: Wannier localization for evaluating integrals involving the screened Coulomb potential and KS orbitals~\cite{gygi2003computation,yu2024gpu}, an inexact Krylov solver for the $Z$-vector equation~\cite{simoncini2003theory,vandeneshof2004inexact,jin2023excited}, and adaptively compressed exchange for hybrid-functional starting points~\cite{lin2016adaptively,yu2025many}. Together with massively parallel GPU implementations of the PDEP construction and hybrid-functional response solver~\cite{yu2022gpu,jin2023excited}, these developments enable BSE forces calculations in supercells containing hundreds of atoms at a cost only a few times larger than that of the corresponding TDDFT forces calculations for hybrid-functional starting points (see Table~SIII for timing details~\cite{SM}).

\begin{figure}[htp!]
    \centering
    \includegraphics[width=\linewidth]{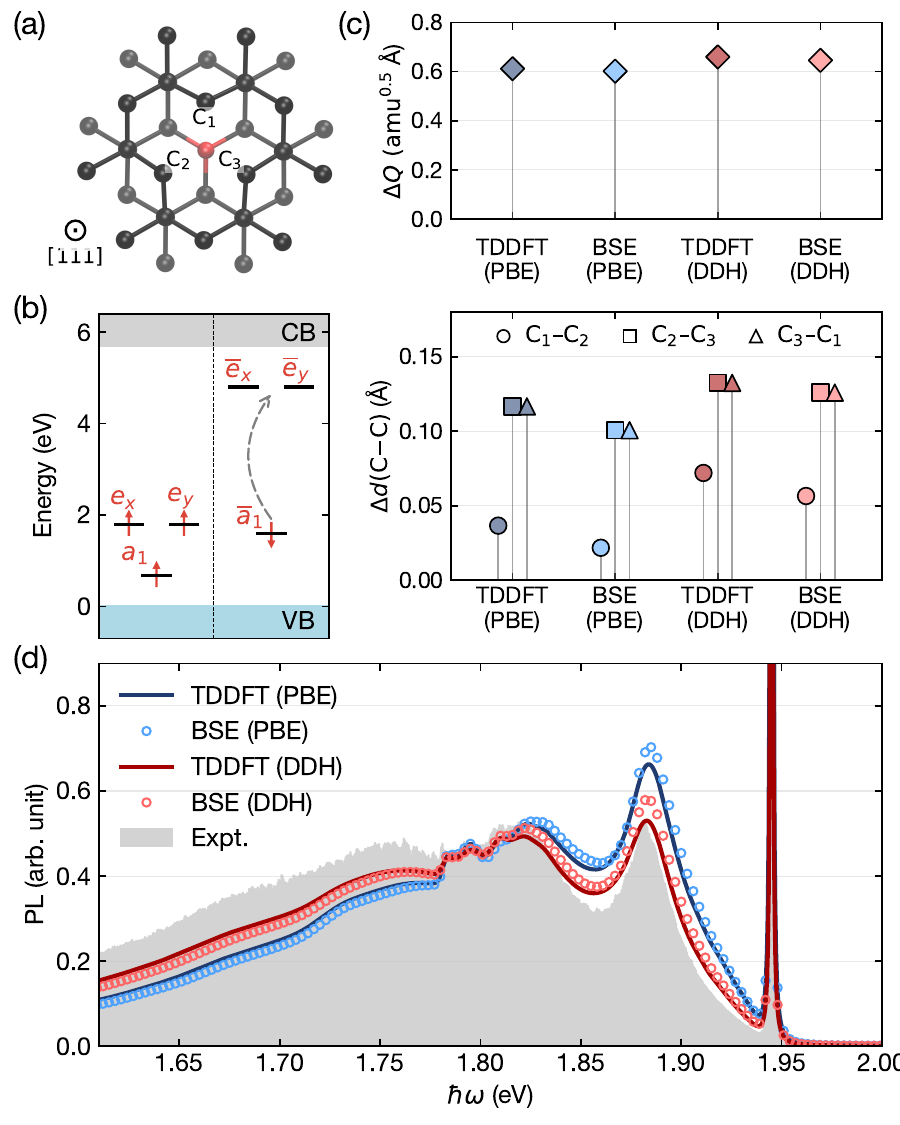}
    \caption{Excited-state relaxation of NV$^-$ in diamond. (a) Atomic structure around the defect. (b) Defect-level diagram and the dominant transition for the ${}^3\!E$ excited state. (c) Total mass-weighted displacement $\Delta Q$ and changes in C--C distances adjacent to the vacancy. (d) Photoluminescence (PL) spectra compared with experiment~\cite{alkauskas2014first}, with the computed line shape shifted to match the measured zero-phonon line (ZPL).}
    \label{fig:nv}
\end{figure}

{\textit{Results}}---We first consider the triplet excited state ${}^3\!E$ of NV$^-$ in diamond, a prototypical spin defect for quantum technologies~\cite{doherty2013nitrogen,gali2019ab,schirhagl2014nitrogen,barry2020sensitivity,childress2013diamond,weber2010quantum,waldherr2014quantum}. This excitation is dominated by a transition from the $a_1$ to the $e_x$ or $e_y$ defect orbitals in the spin-down channel, as illustrated in Fig.~\ref{fig:nv}(b). We relax the ES geometry using analytical BSE forces with DFT starting points obtained from either PBE~\cite{perdew1996generalized} or a dielectric-dependent hybrid (DDH) functional~\cite{skone2014self,skone2016nonempirical}, in which the fraction of exact exchange, 0.18, is set to the inverse of the macroscopic dielectric constant of diamond. The calculations are performed with periodic boundary conditions in a 511-atom supercell, and TDDFT calculations are carried out for comparison. Additional computational details are given in Sec.~SIV.

The vertical excitation energies (VEEs) and adiabatic excitation energies (AEEs) obtained from TDDFT and BSE are summarized in Table~\ref{tab:vee_aee_nv}. For a given starting point, the TDDFT and BSE excitation energies differ by less than 0.1~eV, indicating that TDDFT already provides a reasonable description of this localized defect excitation. The BSE excitation energies reported here are obtained using a fixed scissor operator chosen to reproduce the VEE from the BSE-$G_0W_0$ calculation at the GS geometry.

To quantify ES geometry relaxation, we compute the total mass-weighted displacement between the optimized ES and GS geometries, $\Delta Q = \sqrt{\sum_{I,\alpha} M_I ( R_{I\alpha}^{\mathrm{ES}} - R_{I\alpha}^{\mathrm{GS}} )^2}$, and the changes in the distances between the three carbon atoms adjacent to the vacancy, $\Delta d(\text{C--C})$, as shown in Fig.~\ref{fig:nv}(c). For both PBE and DDH starting points, TDDFT and BSE predict similar values of $\Delta Q$ and $\Delta d(\text{C--C})$, with BSE giving slightly smaller relaxations. This agreement likely reflects the localized character of the NV$^-$ defect orbitals, their energetic separation from the band edges, and the relatively homogeneous dielectric environment of the diamond host. The DDH-based values of $\Delta Q$ are larger than the PBE-based values, consistent with stronger localization of the defect orbitals upon inclusion of exact exchange.

We further compare the relaxed geometries by computing vibrationally resolved photoluminescence (PL) spectra, which encode the projection of the ES structural relaxation onto the vibrational modes of the supercell. As shown in Fig.~\ref{fig:nv}(d), TDDFT and BSE yield similar PL line shapes for both PBE and DDH starting points, and all spectra agree reasonably well with experiment after alignment to the measured zero-phonon line (ZPL). The DDH-based spectra agree better with experiment than the PBE-based spectra, whereas BSE-PBE and TDDFT-PBE perform similarly. These findings indicate that, for NV$^-$ in diamond, the DFT starting point has a larger impact on ES relaxation and vibronic line shapes than the choice between TDDFT and BSE. This sensitivity likely originates from the fact that the single-particle orbitals entering the BSE are obtained from DFT calculations and are not further optimized when computing forces with the BSE. Overall, our results show that, for localized deep-level defects in quasi-homogeneous hosts, BSE and TDDFT can yield similar ES relaxations when based on the same single-particle starting point.

\begin{table}
\caption{Vertical excitation energies (VEEs) and adiabatic excitation energies (AEEs) for NV$^-$ in diamond (eV).}
\label{tab:vee_aee_nv}
\centering
\begin{tabular*}{\columnwidth}{@{\extracolsep{\fill}}lcccccc}
\hline \hline
 & \multicolumn{4}{c}{This work} & \multicolumn{2}{c}{Literature} \\
 & \multicolumn{2}{c}{TDDFT} & \multicolumn{2}{c}{BSE} & Theo. & Expt. \\
 & PBE & DDH & PBE & DDH & & \\ [2pt]
\hline \\ [-8pt]
VEE & 2.09 & 2.40 & 2.17 & 2.43 & 2.24\textsuperscript{a} & \\
AEE & 1.90 & 2.10 & 1.98 & 2.17 & 1.96\textsuperscript{b} & 1.945\textsuperscript{c} \\

\hline \hline
\end{tabular*}

\vspace{1mm}
\begin{minipage}{\columnwidth}
\raggedright
\hspace{0mm}\textsuperscript{a} QDET (DDH)~\cite{chen2025advances}. \hspace{0mm}\textsuperscript{b} $\Delta$SCF (HSE06)~\cite{gali2009theory}. \hspace{0mm}\textsuperscript{c} ZPL~\cite{davies1976optical}. 
\end{minipage}
\end{table}

\begin{figure}[htp!]
    \centering
    \includegraphics[width=\linewidth]{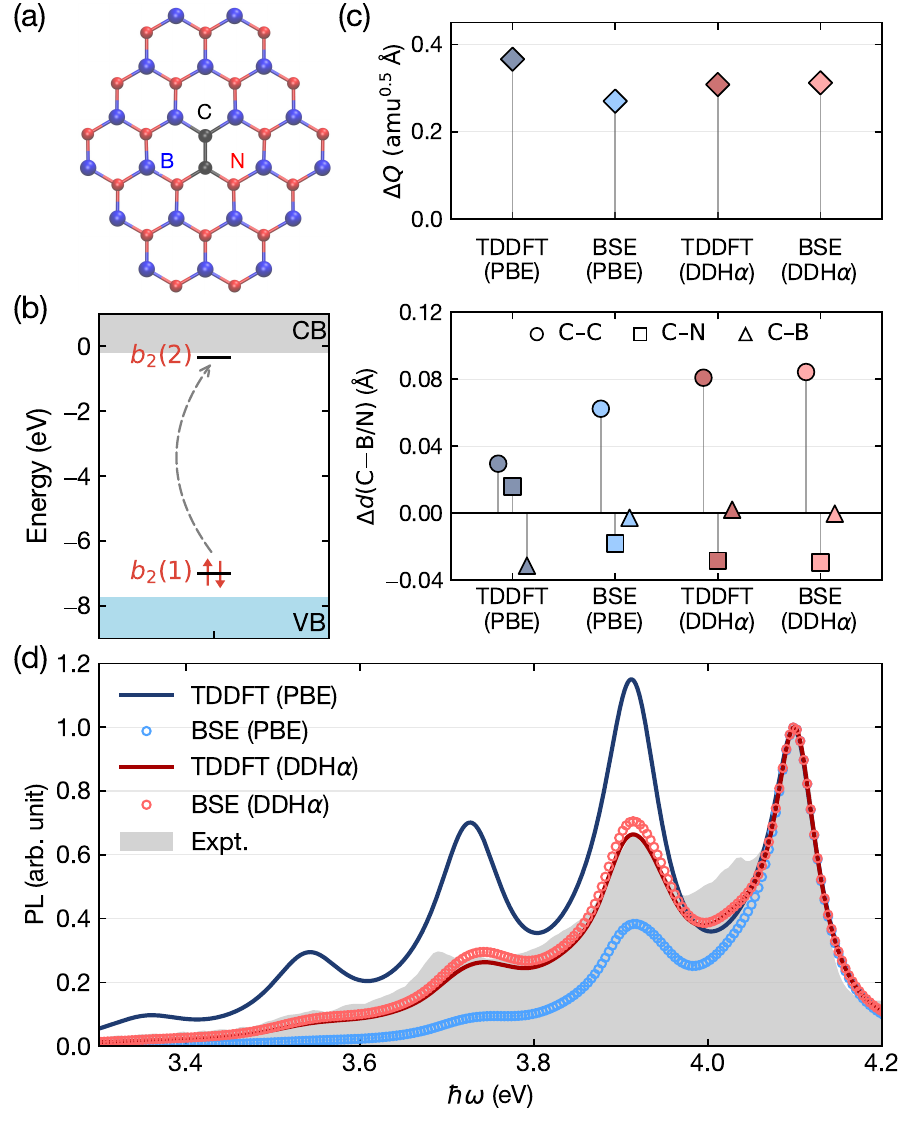}
    \caption{Excited-state relaxation of C$_{\mathrm{B}}$C$_{\mathrm{N}}^0$ in 2D-hBN. (a) Atomic structure of the defect. (b) Defect-level diagram and the dominant transition for the singlet excited state. (c) Total mass-weighted displacement $\Delta Q$ and changes in bond lengths. (d) PL spectra compared with experiment~\cite{bourrellier2016bright}, with the computed line shape shifted to match the measured ZPL.
    }
    \label{fig:cbcn}
\end{figure}

We now turn to C$_\mathrm{B}$C$_\mathrm{N}^{0}$ in 2D-hBN, a defect embedded in a spatially inhomogeneous dielectric environment. This defect has been proposed as a candidate for ultraviolet emitters observed experimentally~\cite{museur2008defect,bourrellier2016bright,mackoit2019carbon}. We compute excitation energies and structural relaxations for the first singlet excited state, dominated by the $b_2(1) \rightarrow b_2(2)$ transition shown in Fig.~\ref{fig:cbcn}(b), using BSE and TDDFT with PBE and DDH$\alpha$ starting points. Because the macroscopic dielectric constant is ill-defined for this 2D system, DDH$\alpha$ is defined here as a PBE0-like functional with an exact-exchange fraction of 0.41, determined by enforcing the generalized Koopmans' condition~\cite{smart2018fundamental}. The calculations are performed in a 288-atom supercell with a vacuum spacing of 20~\AA\ (see Sec.~SIV~\cite{SM} for details).

The computed VEEs and AEEs are summarized in Table~\ref{tab:vee_aee_cbcn}. When using TDDFT, the DDH$\alpha$ excitation energies are about 1 eV larger than the PBE values, a significantly larger difference than that obtained in the case of the NV$^-$ in diamond, highlighting the inaccuracy of semi-local functionals in describing this 2D defect. Even at the BSE level, the DDH$\alpha$ values remain about 0.5 eV larger than the PBE ones, indicating again a non-negligible dependence of the results on the chosen starting point. While BSE with the PBE starting point underestimates the excitation energies, BSE with the DDH$\alpha$ starting point agrees closely (within 0.1 eV) with experimental values and previous theoretical results using cluster models without periodic boundary conditions~\cite{winter2021photoluminescent}.

We next compare the relaxed ES geometries and PL spectra in Fig.~\ref{fig:cbcn}(c,d). TDDFT-PBE predicts a qualitatively different relaxation pattern, with a much larger total mass-weighted displacement $\Delta Q$ and substantially different C--C, C--B, and C--N bond-length changes. This behavior originates from a change in the character of the relaxed excited state: without a screened electron-hole attraction, the TDDFT-PBE state acquires significant valence-band-to-$b_2(2)$ character instead of remaining a localized defect excitation (see Sec.~SV~\cite{SM}). By explicitly including the screened electron-hole attraction, BSE-PBE stabilizes the localized defect excitation and brings the ES geometry closer to the DDH$\alpha$-based results, although the displacement remains underestimated, consistent with the slightly more delocalized excited state at the BSE-PBE level (see Sec.~SV~\cite{SM}). These trends are reflected in the PL spectra: TDDFT-PBE strongly overestimates the phonon sideband, BSE-PBE yields a weaker sideband due to its smaller ES displacement but corrects the qualitative failure of TDDFT-PBE, and TDDFT-DDH$\alpha$ and BSE-DDH$\alpha$ give similar line shapes in close agreement with experiment~\cite{bourrellier2016bright}. Thus, while quantitative excitation energies require the combined treatment of QP corrections and screened electron-hole interactions, as achieved here with BSE-DDH$\alpha$, the ES geometry and vibronic line shape can be described reasonably well by TDDFT when a hybrid-functional starting point provides an adequate single-particle spectrum. More refined hybrid functionals, such as screened-exchange DDH functionals based on local dielectric functions~\cite{zhan2023nonempirical,zhan2025dielectric}, may further improve the ability of TDDFT to describe both ES geometries and excitation energies accurately, as supported by a recent study~\cite{zhan2026hybrid} that predicted a VEE of 4.71 eV and an AEE of 4.59 eV, in close agreement with our BSE-DDH$\alpha$ results.

\begin{table}
\caption{Vertical excitation energies (VEEs) and adiabatic excitation energies (AEEs) for  C$_{\mathrm{B}}$C$_{\mathrm{N}}^0$ in 2D-hBN (eV).}
\label{tab:vee_aee_cbcn}
\centering
\begin{tabular*}{\columnwidth}{@{\extracolsep{\fill}}lcccccc}
\hline \hline
 & \multicolumn{4}{c}{This work} & \multicolumn{2}{c}{Literature} \\
 & \multicolumn{2}{c}{TDDFT} & \multicolumn{2}{c}{BSE} & Theo. & Expt. \\
 & PBE & DDH$\alpha$ & PBE & DDH$\alpha$ & & \\ [2pt]
\hline \\ [-8pt]

VEE & 4.00 & 5.02 & 4.17 & 4.72 & 4.64\textsuperscript{a} & \\
AEE & 3.78 & 4.86 & 4.09 & 4.57 & 4.49\textsuperscript{a} & 4.5\textsuperscript{b} \\
\hline \hline
\end{tabular*}

\vspace{1mm}
\begin{minipage}{\columnwidth}
\raggedright

\hspace{0mm}\textsuperscript{a} VEE obtained using BSE-ev$GW$ (PBE0) and ES geometry relaxed using TDDFT (PBE0) forces~\cite{winter2021photoluminescent}.

\hspace{0mm}\textsuperscript{b} Estimated as the experimental ZPL of 4.08 eV~\cite{museur2008defect} with a 0.13 eV zero-point vibrational energy contribution and 0.3 eV multilayer polarization effects contribution~\cite{winter2021photoluminescent}. 
\end{minipage}
\end{table}

\begin{figure}
    \centering
    \includegraphics[width=0.9\linewidth]{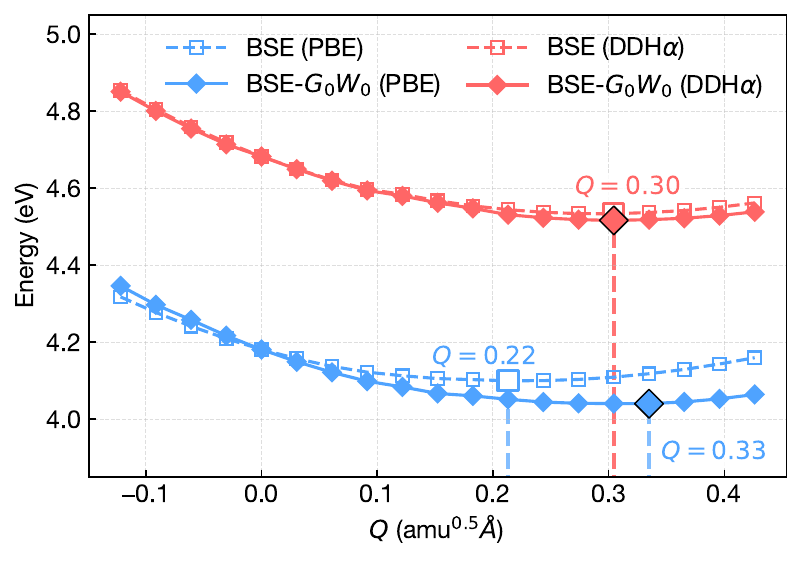}
    \caption{Potential energy curves computed with different methods for C$_\mathrm{B}$C$_\mathrm{N}^{0}$ in 2D-hBN along the configuration coordinate connecting the GS and ES geometries. Highlighted larger markers indicate local minima along the path.}
    \label{fig:cbcn_pes}
\end{figure}

Finally, we assess the reliability of the fixed-scissor approximation used in the BSE forces calculations by computing BSE potential energy curves for C$_\mathrm{B}$C$_\mathrm{N}^{0}$ along the mass-weighted interpolation coordinate connecting the GS and BSE-DDH$\alpha$ optimized ES geometries, using a 128-atom supercell. As shown in Fig.~\ref{fig:cbcn_pes}, fixed-scissor BSE and BSE-$G_0W_0$ give consistent PECs for the DDH$\alpha$ starting point, with both curves exhibiting a local minimum near the BSE-DDH$\alpha$ relaxed geometry. In contrast, for the PBE starting point, the fixed-scissor and BSE-$G_0W_0$ PECs differ more substantially, indicating that the QP correction is large and geometry-dependent. This behavior is consistent with the fact that the PBE starting point lacks exact exchange and provides an inadequate single-particle spectrum for the localized defect excitation, and thus the QP correction cannot be treated as a simple geometry-independent scissor shift. The DDH$\alpha$ starting point partially incorporates geometry-dependent QP effects through exact exchange, reducing the remaining self-energy correction and making the fixed-scissor approximation more reliable. These comparisons indicate that a hybrid-functional starting point can reduce the magnitude and geometry dependence of the remaining QP correction, thereby making the fixed-scissor approximation reliable for BSE analytical forces.

{\textit{Discussion}}--- We have developed and validated an efficient plane-wave implementation of analytical BSE forces for ES calculations in complex materials. Our approach removes the principal computational barriers that have historically limited BSE-level ES structural relaxation studies to small molecules. Specifically, the combination of a $Z$-vector Lagrangian formulation, a DMPT approach, and a low-rank representation of the dielectric matrix yields a favorable scaling. Together with GPU-accelerated parallelism, our approach enables the ES structural relaxation in supercells containing several hundred atoms for the first time.

Our applications identify when BSE forces provide information beyond TDDFT and when TDDFT may already be sufficient. For NV$^-$ in diamond, a localized deep-level defect in a quasi-homogeneous host, TDDFT and BSE predict quantitatively similar ES displacements and PL spectra when based on the same single-particle starting point. This agreement suggests that hybrid-functional TDDFT can be a reliable and cost-effective surrogate for ES structural relaxation in this class of systems. In contrast, for C$_{\mathrm{B}}$C$_\mathrm{N}^0$ in 2D-hBN, the spatially inhomogeneous dielectric environment makes the explicit treatment of screened electron-hole interaction more important. For this system, BSE with a DDH$\alpha$ functional yields accurate excitation energies, robust ES geometries, and PL spectra in close agreement with experiment. Instead, semi-local TDDFT qualitatively fails by relaxing toward an excited state with incorrect character -- a finding with direct implications for computational screening of quantum emitters in 2D materials~\cite{ivady2020ab,lee2022spin,bertoldo2022quantum}. The approach developed here opens the door to BSE-level studies of ES relaxation in systems where electron-hole interactions, dielectric inhomogeneity, and lattice relaxation are strongly coupled, including self-trapped excitons in metal-halide perovskites~\cite{li2019self,li2025luminescence,sun2023toward,jin2024self} and defects in oxides~\cite{davidsson2024discovery,somjit2025nv,zhang2026deep}.

Several methodological extensions remain important. Lifting the fixed-scissor approximation through geometry-dependent QP corrections may improve the accuracy of the approach for systems with strong self-energy variations along the relaxation path. More generally, adopting a self-consistent $GW$ workflow would reduce the explicit starting-point dependence of the BSE forces calculations. Further, ongoing efforts to formulate analytical non-adiabatic couplings within the present Lagrangian framework would enable both investigations of exciton-phonon coupling and dynamical simulations of non-radiative decay~\cite{villani2026first}. Finally, integration with machine-learned force fields~\cite{malosso2026transferable} or surrogate dielectric models~\cite{dong2021machine,zauchner2023accelerating} could further extend accessible timescales and system sizes. All these directions position the present implementation as a foundation for predictive, BSE-level studies of ES phenomena across a wide class of materials.

\textit{Codes used}---\textsc{Quantum ESPRESSO}~\cite{giannozzi2020qe,carnimeo2023quantum} is used for DFT calculations, \textsc{WEST}~\cite{govoni2015large} is used for BSE and TDDFT calculations, \textsc{PyPL}~\cite{pypl} is used for PL spectra calculations.

\textit{Acknowledgments}---The theoretical and computational work was primarily supported by the Midwest Integrated Center for Computational Materials (MICCoM) as part of the Computational Materials Sciences Program funded by the U.S. Department of Energy. This research used resources of the National Energy Research Scientific Computing Center (NERSC), a DOE Office of Science User Facility supported by the Office of Science of the U.S. Department of Energy under contract No.~DE-AC02-05CH11231 using NERSC Award No.~ALCC-ERCAP0025950, and resources of the University of Chicago Research Computing Center. The Flatiron Institute is a division of the Simons Foundation.

\textit{Data availability}---The data that support the findings of this Letter are openly available~\cite{data_statement}.


\nocite{budzak2017accurate,adamo1999toward,heyd2003hybrid,krukau2006influence,hamann2013optimized,schlipf2015optimization,phonopy-phono3py-JPCM,phonopy-phono3py-JPSJ,razinkovas2021vibrational,jin2021photoluminescence,muechler2022quantum,haldar2023local,benedek2025accurate,martirez2026optical,lau2024optical,korona2019exploring,li2024accurate,li2024particle}
\bibliography{Main}

\appendix
\section*{End Matter}

\begin{figure}[!htp]
\centering
\includegraphics[width=\linewidth]{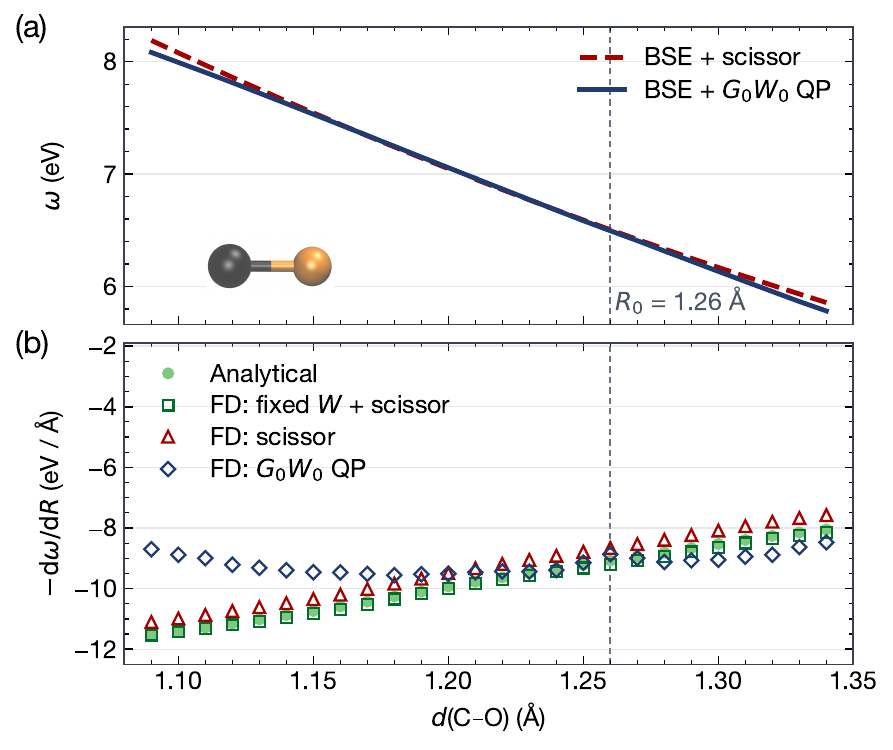}
\caption{Validation of BSE forces for the lowest singlet excited state of carbon monoxide. (a) BSE vertical excitation energies obtained using $G_0W_0$ quasiparticle (QP) energies and using a scissor operator. (b) Analytical excitation energy derivative compared with finite-difference (FD) derivative obtained with different approximations to QP corrections and $W$. The vertical dashed line marks the ES bond length.}
\label{fig:co}
\end{figure}

We validate the approximations used in the present BSE analytical forces implementation for the lowest singlet excited state of carbon monoxide, as shown in Fig.~\ref{fig:co}. The BSE vertical excitation energies obtained using explicit $G_0W_0$ QP energies and a scissor operator fitted at the GS bond length agree closely over the bond-length range considered here. The analytical derivative agrees precisely with the finite-difference (FD) derivative evaluated at the same level of approximation, namely with fixed $W$ and a fixed scissor operator. Updating $W$ at each bond length while keeping the scissor operator fixed leads to only small changes, indicating that neglecting $\delta W$ has a minor effect in this case. FD derivatives computed with explicit $G_0W_0$ QP energies and updated $W$ agree closely with the analytical result near the relaxed ES bond length, although larger deviations appear at shorter bond lengths. These results show that the accuracy of the fixed-scissor approximation is geometry-dependent and that it remains reliable in the region relevant to the relaxed excited state. Going beyond this approximation requires a fully self-consistent treatment of QP-energy derivatives and is left for future work.

\end{document}



\title{
Supplemental Material for ``Analytical Forces from the Bethe-Salpeter Equation for Large-Scale Excited-State Relaxation"
}

\author{Yu Jin}
\email{yjin@flatironinstitute.org}
\affiliation{Pritzker School of Molecular Engineering, University of Chicago, Chicago, Illinois 60637, USA}
\affiliation{Initiative for Computational Catalysis, Flatiron Institute, New York, New York 10010, USA}
\author{Victor Wen-zhe Yu}
\affiliation{Materials Science Division, Argonne National Laboratory, Lemont, Illinois 60439, USA}
\author{Marco Govoni}
\affiliation{Department of Physics, Computer Science, and Mathematics, University of Modena and Reggio Emilia, Modena, 41125, Italy}
\author{Giulia Galli}
\email{gagalli@uchicago.edu}
\affiliation{Pritzker School of Molecular Engineering, University of Chicago, Chicago, Illinois 60637, USA}
\affiliation{Department of Chemistry, University of Chicago, Chicago, Illinois 60637, USA}
\affiliation{Materials Science Division, Argonne National Laboratory, Lemont, Illinois 60439, USA}

\date{\today}

\maketitle

\tableofcontents
\newpage

\section{Additional equations for analytical BSE forces}

Equation~(1) of the main text gives the Bethe-Salpeter equation (BSE) in the Tamm-Dancoff approximation (TDA), which is the form used in this work. For completeness, we first present the corresponding BSE beyond the TDA~\cite{nguyen2019finite}:
\begin{equation}
    \left( \begin{matrix}
        \mathcal{D} + 2\mathcal{K}^{\mathrm{1e}} - \mathcal{K}^{\mathrm{1d}} & 2\mathcal{K}^{\mathrm{2e}} - \mathcal{K}^{\mathrm{2d}} \\
        2\mathcal{K}^{\mathrm{2e}*} - \mathcal{K}^{\mathrm{2d}*} & \mathcal{D}^* + 2\mathcal{K}^{\mathrm{1e}*} - \mathcal{K}^{\mathrm{1d}*} \\
    \end{matrix} \right) \left( \begin{matrix} \mathcal{A}_s \\ \mathcal{B}_s \end{matrix} \right) = \omega_s \left( \begin{matrix}
        \mathcal{I} & 0 \\ 0 & - \mathcal{I}
    \end{matrix}  \right) \left( \begin{matrix}
        \mathcal{A}_s \\ \mathcal{B}_s
    \end{matrix} \right).
\end{equation}
Here, the response operators $\mathcal{D}$, $\mathcal{K}^{\mathrm{1e}}$, and $\mathcal{K}^{\mathrm{1d}}$ are defined in Eq.~(2) of the main text. Beyond the TDA, two additional coupling operators enter:
\begin{subequations}
\begin{align}
    \left( \mathcal{K}^{\mathrm{2e}}\mathcal{A} \right)_v &= \hat{P}_c \int d\mathbf{r}' v_c (\mathbf{r},\mathbf{r}') \sum_{v'} a_{v'}^{\ast}(\mathbf{r}') \varphi_{v'}(\mathbf{r}') |\varphi_v\rangle, \\
    \left( \mathcal{K}^{\mathrm{2d}}\mathcal{A} \right)_v &= \hat{P}_c \sum_{v'} \int d\mathbf{r}' W (\mathbf{r},\mathbf{r}') a_{v'}^{\ast}(\mathbf{r}') \varphi_v(\mathbf{r}') |\varphi_{v'}\rangle .
\end{align}
\end{subequations}
The set $\mathcal{B}_s \equiv { |b_{s,v}\rangle, v=1,\ldots,N_{\mathrm{occ}}}$, together with $\mathcal{A}_s$, defines the transition density matrix as
\begin{equation}
    \Delta \hat{\rho} =\sum_{v=1}^{N_{\mathrm{occ}}} \left( | a_{v} \rangle \langle \varphi_v | + | \varphi_v \rangle \langle b_v | \right).
\end{equation}
Within the TDA, $\mathcal{B}_s$ is set to zero, and only the terms involving $\mathcal{A}_s$ are retained. This is the approximation adopted in the present implementation.

Within the TDA, the analytical derivative of the excitation energy $\omega$ can be evaluated efficiently by introducing an extended Lagrangian~\cite{hutter2003excited,zhang2015subspace,jin2023excited}:
\begin{equation}
\begin{aligned}
    \mathcal{L} \left[ {|\varphi_v\rangle}, \mathcal{A}, \omega, \Lambda, {|Z_v\rangle} \right] &= \mathcal{A}^{\dagger} \left( \mathcal{D} + 2\mathcal{K}^{\mathrm{1e}} - \mathcal{K}^{\mathrm{1d}} \right) \mathcal{A} - \omega \left( \mathcal{A}^{\dagger}\mathcal{A}-1 \right) + \sum_v \big\langle Z_v | \Big( \hat{H}^{\mathrm{KS}}|\varphi_v\rangle - \sum_{v'} \Lambda_{vv'}|\varphi_{v'}\rangle \Big).
\end{aligned}
\label{eq:L}
\end{equation}
The stationarity conditions of the Lagrangian recover the equations needed for the analytical-derivative formalism. Stationarity with respect to $\mathcal{A}$ gives the BSE in Eq.~(1) of the main text, while stationarity with respect to $\omega$ imposes the normalization condition $\mathcal{A}^{\dagger}\mathcal{A}=1$. Stationarity with respect to $\Lambda$ constrains the $Z$-vectors, $\mathcal{Z}=\{|Z_v\rangle, v=1,\ldots,N_{\mathrm{occ}}\}$, to lie in the unoccupied subspace, i.e., $\langle Z_v|\varphi_{v'}\rangle=0$. Stationarity with respect to $|Z_v\rangle$ yields the Kohn-Sham (KS) equations. The remaining stationarity condition, $\delta \mathcal{L}/\delta \langle \varphi_v|=0$, gives the $Z$-vector equation reported in Eq.~(7) of the main text. Once these stationarity conditions are satisfied, the total derivative of $\omega$ with respect to a nuclear coordinate $R$ is obtained from the explicit derivative of the Lagrangian:
\begin{equation}
    \frac{d\omega}{dR} = \frac{\partial \mathcal{L}}{\partial R}.
\end{equation}

We next define the KS response operators that appear in the $Z$-vector equation, Eq.~(7) of the main text:
\begin{subequations}
\begin{align}
    \left( \mathcal{D}^{\mathrm{KS}}\mathcal{A} \right)_v &= \hat{P}_c \left( \hat{H}^{\mathrm{KS}} - \varepsilon_v^{\mathrm{KS}} \right) | a_v \rangle \\
    \left( \mathcal{K}^{\mathrm{1e},\mathrm{KS}} \mathcal{A} \right)_v &= \hat{P}_c \int d\mathbf{r}' f_{\mathrm{Hxc}}^{\mathrm{loc}}(\mathbf{r},\mathbf{r}') \sum_{v'} \varphi_{v'}^{\ast}(\mathbf{r}') a_{v'}(\mathbf{r}') |\varphi_v\rangle , \\
    \left( \mathcal{K}^{\mathrm{1d},\mathrm{KS}} \mathcal{A} \right)_v &= \alpha_{\mathrm{EXX}} \hat{P}_c \sum_{v'} \int d\mathbf{r}' v_c(\mathbf{r},\mathbf{r}') \varphi_{v'}^{\ast}(\mathbf{r}') \varphi_v(\mathbf{r}') |a_{v'}\rangle , \\
    \left( \mathcal{K}^{\mathrm{2e},\mathrm{KS}} \mathcal{A} \right)_v &= \hat{P}_c \int d\mathbf{r}' f_{\mathrm{Hxc}}^{\mathrm{loc}}(\mathbf{r},\mathbf{r}') \sum_{v'} a_{v'}^{\ast}(\mathbf{r}') \varphi_{v'}(\mathbf{r}') |\varphi_v\rangle, \\
    \left( \mathcal{K}^{\mathrm{2d},{\mathrm{KS}}} \mathcal{A} \right)_v &= \alpha_{\mathrm{EXX}} \hat{P}_c \sum_{v'} \int d\mathbf{r}' v_c(\mathbf{r},\mathbf{r}') a_{v'}^{\ast}(\mathbf{r}') \varphi_v(\mathbf{r}') |\varphi_{v'}\rangle .
\end{align}
\end{subequations}
Here, $f_{\mathrm{Hxc}}^{\mathrm{loc}}=v_c+f_{\mathrm{xc}}^{\mathrm{loc}}$, where $f_{\mathrm{xc}}^{\mathrm{loc}}$ denotes the local part of the exchange-correlation kernel. The parameter $\alpha_{\mathrm{EXX}}$ is the fraction of exact exchange and is zero for local and semilocal DFT starting points. These KS response operators enter the $Z$-vector equation because the single-particle orbitals used in the BSE are obtained from the underlying KS calculation and are not further optimized at the BSE level. They are also the response operators appearing in the corresponding TDDFT eigenvalue problem. A detailed comparison between the TDDFT and BSE response operators is given in Table~\ref{s-tbl:tddft_bse_comparison}.

The vector $\mathcal{U}$ on the right-hand side of Eq.~(7) of the main text is obtained by differentiating the TDA BSE energy expression with respect to the occupied KS orbitals:
\begin{equation}
    |u_v\rangle = \hat{P}_c \frac{ \delta \left[ \mathcal{A}^{\dagger} \left( \mathcal{D} + 2 \mathcal{K}^{\text{1e}} - \mathcal{K}^{\text{1d}} \right) \mathcal{A} \right]}{\delta \langle \varphi_v |}.
\label{s-eq:u}
\end{equation}
The first contribution to Eq.~\eqref{s-eq:u} arises from the quasiparticle energy-difference operator $\mathcal{D}$:
\begin{equation}
\begin{aligned}
    \hat{P}_c \frac{ \delta \left( \mathcal{A}^{\dagger} \mathcal{D} \mathcal{A} \right)}{\delta \langle \varphi_v|} &= \hat{P}_c \frac{\delta \left[ \sum_{v'v''} \left\langle a_{v'} \left| \hat{P}_c \left( \hat{H}^{\mathrm{QP}}\delta_{v'v''} - \varepsilon_{v'v''}^{\mathrm{QP}} \right) \right| a_{v''}\right\rangle \right]}{\delta \langle \varphi_v|}.
\end{aligned}
\end{equation}
Using the scissor-operator approximation for the quasiparticle correction, $\hat{H}^{\mathrm{QP}}\approx \hat{H}^{\mathrm{KS}}+\hat{P}_c \Delta_{\mathrm{sci.}}$, we obtain
\begin{equation}
\begin{aligned}
    \hat{P}_c \frac{ \delta \left(\mathcal{A}^{\dagger} \mathcal{D} \mathcal{A} \right)}{\delta \langle \varphi_v|} &\approx \hat{P}_c \frac{ \delta \left[ \sum_{v'v''} \left\langle a_{v'} \left| \hat{P}_c \left( \hat{H}^{\mathrm{KS}}\delta_{v'v''} + \hat{P}_c\Delta_{\mathrm{sci.}}\delta_{v'v''} - \left\langle \varphi_{v'} \left| \hat{H}^{\mathrm{KS}} + \hat{P}_c \Delta_{\mathrm{sci.}} \right| \varphi_{v''}\right\rangle \right) \right| a_{v''}\right\rangle \right]}{\delta \langle \varphi_v|} \\
    &= \hat{P}_c \frac{ \delta \left[ \sum_{v'v''} \left\langle a_{v'} \left| \hat{P}_c \left( \hat{H}^{\mathrm{KS}}\delta_{v'v''} - \left\langle \varphi_{v'} \left| \hat{H}^{\mathrm{KS}} \right| \varphi_{v''}\right\rangle \right) \right| a_{v''}\right\rangle + \Delta_{\mathrm{sci.}} \right]}{\delta \langle \varphi_v|} \\ 
    &= \hat{P}_c \int d\mathbf{r}' f_{\mathrm{Hxc}}^{\mathrm{loc}}(\mathbf{r},\mathbf{r}') \Delta\rho^{(x)}(\mathbf{r}') |\varphi_v\rangle - \alpha_{\mathrm{EXX}} \hat{P}_c \sum_{v'=1}^{N_{\mathrm{occ}}} \int d\mathbf{r}' v_c(\mathbf{r},\mathbf{r}') a_{v'}^{\ast}(\mathbf{r}') \varphi_v(\mathbf{r}') |a_{v'}\rangle \\
    &\quad + \alpha_{\mathrm{EXX}} \hat{P}_c \sum_{v'=1}^{N_{\mathrm{occ}}} \int d\mathbf{r}' v_c(\mathbf{r},\mathbf{r}') \varphi_{v'}^{\ast}(\mathbf{r}') \varphi_v(\mathbf{r}') \sum_{v''=1}^{N_{\mathrm{occ}}} \langle a_{v''}|a_{v'}\rangle |\varphi_{v''}\rangle .
\end{aligned}
\end{equation}
This contribution is therefore identical to that obtained in LR-TDDFT~\cite{jin2023excited}. Going beyond the scissor-operator approximation would require evaluating derivatives of the frequency-dependent self-energy operator,
$\delta[\hat{\Sigma}(\varepsilon^{\mathrm{QP}})]/\delta \langle \varphi_v|$,
which is an important direction for future work.

The second contribution to Eq.~\eqref{s-eq:u} comes from the bare exchange-like electron-hole interaction, $2\mathcal{K}^{\mathrm{1e}}$:
\begin{equation}
\begin{aligned}
    \hat{P}_c \frac{ \delta \left( 2\mathcal{A}^{\dagger} \mathcal{K}^{\mathrm{1e}} \mathcal{A} \right)}{\delta \langle \varphi_v|} &= 2\hat{P}_c \int d\mathbf{r}' v_c(\mathbf{r},\mathbf{r}') \Delta\rho^{\ast}(\mathbf{r}') |a_v\rangle - 2\sum_{v'=1}^{N_{\mathrm{occ}}} \left[ \int d\mathbf{r}' \varphi_{v'}^{\ast}(\mathbf{r}') \varphi_v(\mathbf{r}') \int d\mathbf{r}'' v_c(\mathbf{r}',\mathbf{r}'') \Delta\rho(\mathbf{r}'') \right] |a_{v'}\rangle ,
\end{aligned}
\end{equation}
where
\begin{equation}
    \Delta\rho(\mathbf{r}) = \sum_{v=1}^{N_{\mathrm{occ}}} \varphi_v^{\ast}(\mathbf{r})a_v(\mathbf{r}) .
\end{equation}
is the transition density. This term is evaluated exactly in the present implementation.

The third contribution to Eq.~\eqref{s-eq:u} comes from the screened direct electron-hole interaction, $\mathcal{K}^{\mathrm{1d}}$:
\begin{equation}
\begin{aligned}
    - \hat{P}_c \frac{ \delta \left( \mathcal{A}^{\dagger} \mathcal{K}^{\mathrm{1d}} \mathcal{A} \right) }{\delta \langle \varphi_v|} &= - \hat{P}_c \sum_{v'=1}^{N_{\mathrm{occ}}} \int d\mathbf{r}' W(\mathbf{r},\mathbf{r}') a_{v'}^{\ast}(\mathbf{r}') a_v(\mathbf{r}') |\varphi_{v'}\rangle \\
    &\quad + \sum_{v'=1}^{N_{\mathrm{occ}}} \left[ \int d\mathbf{r}' \sum_{v''=1}^{N_{\mathrm{occ}}} a_{v''}^{\ast}(\mathbf{r}') \varphi_v(\mathbf{r}') \int d\mathbf{r}'' W(\mathbf{r}',\mathbf{r}'') \varphi_{v'}^{\ast}(\mathbf{r}'') \varphi_{v''}(\mathbf{r}'') \right] |a_{v'}\rangle .
\end{aligned}
\end{equation}
In deriving this term, we neglect the derivative of the screened Coulomb interaction with respect to the occupied KS orbitals, i.e.,
$\delta W/\delta \langle \varphi_v| \approx 0$.
This approximation is analogous to the standard approximation used in deriving the BSE, namely $\delta W/\delta G \approx 0$~\cite{rohlfing2000electron,villalobos2023lagrangian}. Its effect is assessed in the main text using carbon monoxide as a test case.

For completeness, Table~\ref{s-tbl:tddft_bse_comparison} compares TDDFT and BSE at the level of the TDA eigenvalue equation. Semi-local TDDFT does not contain a direct electron-hole attraction term. Hybrid TDDFT includes an analogous exchange contribution, in which the screened Coulomb interaction is effectively approximated by the bare Coulomb interaction scaled by the exact-exchange fraction $\alpha_{\mathrm{EXX}}$. Therefore, for systems with relatively homogeneous dielectric screening, a dielectric-dependent hybrid (DDH) functional with $\alpha_{\mathrm{EXX}}=1/\epsilon_{\infty}$ can approximately capture the screened electron-hole interaction~\cite{skone2014self,skone2016nonempirical}. For systems with spatially inhomogeneous dielectric screening, such as low-dimensional materials and interfaces, this approximation is less reliable, motivating either improved hybrid functionals or an explicit BSE treatment.

\begin{table}[H]
    \caption{Comparison between TDDFT and BSE in terms of the TDA eigenvalue problem $\left(\mathcal{D} + 2\mathcal{K}^{\mathrm{1e}} - \mathcal{K}^{\mathrm{1d}} \right)\mathcal{A} = \omega\mathcal{A}$.}
    \label{s-tbl:tddft_bse_comparison}
    \centering
    \renewcommand{\arraystretch}{1.25}
    \begin{tabular}{cccc}
    \hline\hline
    & $(\mathcal{DA})_v$  & $(\mathcal{K}^{\mathrm{1e}}\mathcal{A})_v$  & $(\mathcal{K}^{\mathrm{1d}}\mathcal{A})_v$ \\
    \hline
    TDDFT (semi-local) & $\displaystyle \hat{P}_{c} \left( \hat{H}^{\text{KS}} - \varepsilon_{v}^{\text{KS}} \right) | a_{v} \rangle$ & $\displaystyle \hat{P}_c  \int d \mathbf{r}' f_{\mathrm{Hxc}}^{\mathrm{loc}}(\mathbf{r},\mathbf{r'}) \Delta \rho (\mathbf{r'}) | \varphi_v \rangle$ & 0 \\
    TDDFT (hybrid) & $\displaystyle \hat{P}_{c} \left( \hat{H}^{\text{KS}} - \varepsilon_{v}^{\text{KS}} \right) | a_{v} \rangle$ & $\displaystyle \hat{P}_c  \int d \mathbf{r}' f_{\mathrm{Hxc}}^{\mathrm{loc}}(\mathbf{r},\mathbf{r'}) \Delta \rho (\mathbf{r'}) | \varphi_v \rangle$ & $\displaystyle \alpha_{\mathrm{EXX}} \hat{P}_c \sum_{v'}  
    \int d \mathbf{r'} v_c(\mathbf{r},\mathbf{r'}) \varphi_{v'}^{\ast} (\mathbf{r'}) \varphi_{v} (\mathbf{r'}) | a_{v'} \rangle$   \\
    BSE-Scissor & $\displaystyle \hat{P}_{c} \left( \hat{H}^{\text{KS}} + \hat{P}_{c} \Delta_{\mathrm{sci.}} - \varepsilon_{v}^{\text{KS}} \right) | a_{v} \rangle$ & $\displaystyle \hat{P}_c  \int d \mathbf{r}' v_c(\mathbf{r},\mathbf{r'}) \Delta \rho (\mathbf{r'}) | \varphi_v \rangle$ & $\displaystyle \hat{P}_c \sum_{v'}  
    \int d \mathbf{r'} W(\mathbf{r},\mathbf{r'}) \varphi_{v'}^{\ast} (\mathbf{r'}) \varphi_{v} (\mathbf{r'}) | a_{v'} \rangle$  \\
    BSE-$G_0W_0$ & $\displaystyle \hat{P}_{c} \left( \hat{H}^{\text{QP}} - \varepsilon_{v}^{\text{QP}} \right) | a_{v} \rangle$ & $\displaystyle \hat{P}_c  \int d \mathbf{r}' v_c(\mathbf{r},\mathbf{r'}) \Delta \rho (\mathbf{r'}) | \varphi_v \rangle$ & $\displaystyle \hat{P}_c \sum_{v'}  
    \int d \mathbf{r'} W(\mathbf{r},\mathbf{r'}) \varphi_{v'}^{\ast} (\mathbf{r'}) \varphi_{v} (\mathbf{r'}) | a_{v'} \rangle$ \\
     \hline\hline
   \end{tabular}
\end{table}

\section{Validation of analytical BSE forces}

To validate the implementation of analytical BSE forces, we compare them with numerical finite-difference forces for the ${}^3\!E$ excited state of NV$^-$ in diamond using a 63-atom supercell. The tests are performed with PBE~\cite{perdew1996generalized}, DDH, and HSE~\cite{heyd2003hybrid,krukau2006influence} starting points. We compare the analytical and numerical forces on the three carbon atoms adjacent to the vacancy, labeled C29, C42, and C56, with the results summarized in Table~\ref{s-tbl:nvtriplet}. A displaced NV$^-$ geometry is used to lift the degeneracy of the ${}^3\!E$ excited state.

The numerical forces are computed by finite differences using displacement amplitudes of 0.003568 \AA\ to 0.007136 \AA. To remain consistent with the $\dfrac{\delta W}{\delta \langle \varphi_v |} \approx 0$ approximation used in the analytical BSE forces implementation, the projective dielectric eigenpotentials (PDEPs) used to evaluate $W$ are computed at the central geometry and reused for the forward and backward displaced geometries.

The differences between analytical and numerical forces are negligible for all three starting points, validating the correctness of the analytical BSE forces implementation.

\begin{table}[H]
    \caption{Analytical and numerical finite-difference forces (Ry/bohr) for the vertical excitation energy of the ${}^3\!E$ excited state of NV$^-$ in diamond, computed using BSE with PBE, DDH, and HSE starting points. C29, C42, and C56 denote the three carbon atoms adjacent to the vacancy site.}
   \label{s-tbl:nvtriplet}
   \centering
   \renewcommand{\arraystretch}{1.25}
   {\begin{tabular}{ccccccccccc}
     \hline\hline
     & Atom & \multicolumn{3}{c}{Analytical forces} & \multicolumn{3}{c}{Numerical forces} & \multicolumn{3}{c}{Differences} \\
     & & $X$ & $Y$ & $Z$ & $X$ & $Y$ & $Z$ & $X$ & $Y$ & $Z$ \\
     \hline
     & C29 & \hphantom{$+$}0.025821 & \hphantom{$+$}0.016590 & \hphantom{$+$}0.009515 & \hphantom{$+$}0.025849 & \hphantom{$+$}0.016584 & \hphantom{$+$}0.009516 & $-0.000028$ & \hphantom{$+$}0.000006 & $-0.000001$ \\
     PBE& C42 & \hphantom{$+$}0.016338 & \hphantom{$+$}0.025656 & \hphantom{$+$}0.009410 & \hphantom{$+$}0.016339 & \hphantom{$+$}0.025646 & \hphantom{$+$}0.009417 & $-0.000001$ & \hphantom{$+$}0.000001 & $-0.000007$ \\
     & C56 & \hphantom{$+$}0.004201 & \hphantom{$+$}0.004530 & $-0.041964$ & \hphantom{$+$}0.004212 & \hphantom{$+$}0.004533 & $-0.041955$ & $-0.000011$ & $-0.000003$ & $-0.000009$ \\
     \hline
     & C29 & \hphantom{$+$}0.035369 & \hphantom{$+$}0.011273 & \hphantom{$+$}0.015964 & \hphantom{$+$}0.035419 & \hphantom{$+$}0.011229 & \hphantom{$+$}0.015961 & $-0.000050$ & \hphantom{$+$}0.000044 & \hphantom{$+$}0.000003 \\
     DDH& C42 & \hphantom{$+$}0.010975 & \hphantom{$+$}0.035256 & \hphantom{$+$}0.015869 & \hphantom{$+$}0.010931 & \hphantom{$+$}0.035306 & \hphantom{$+$}0.015867 & \hphantom{$+$}0.000044 & $-0.000050$ & \hphantom{$+$}0.000002 \\
     & C56 & \hphantom{$+$}0.000928 & \hphantom{$+$}0.001329 & $-0.047988$ & \hphantom{$+$}0.000917 & \hphantom{$+$}0.001313 & $-0.047893$ & \hphantom{$+$}0.000011 & \hphantom{$+$}0.000016 & $-0.000095$ \\
     \hline
     & C29 & \hphantom{$+$}0.038963 & \hphantom{$+$}0.009219 & \hphantom{$+$}0.018567 & \hphantom{$+$}0.039019 & \hphantom{$+$}0.009176 & \hphantom{$+$}0.018565 & $-0.000056$ & \hphantom{$+$}0.000043 & \hphantom{$+$}0.000002 \\
     HSE& C42 & \hphantom{$+$}0.008916 & \hphantom{$+$}0.038855 & \hphantom{$+$}0.018463 & \hphantom{$+$}0.008870 & \hphantom{$+$}0.038911 & \hphantom{$+$}0.018461 & \hphantom{$+$}0.000046 & $-0.000056$ & \hphantom{$+$}0.000002 \\
     & C56 & $-0.001507$ & $-0.001097$ & $-0.051321$ & $-0.001512$ & $-0.001102$ & $-0.051212$ & \hphantom{$+$}0.000005 & \hphantom{$+$}0.000005 & $-0.000109$ \\
     \hline\hline
   \end{tabular}}
\end{table}

\newpage
\section{Applications to small molecules}

\begin{figure}[H]
    \centering
    \includegraphics[width=\linewidth]{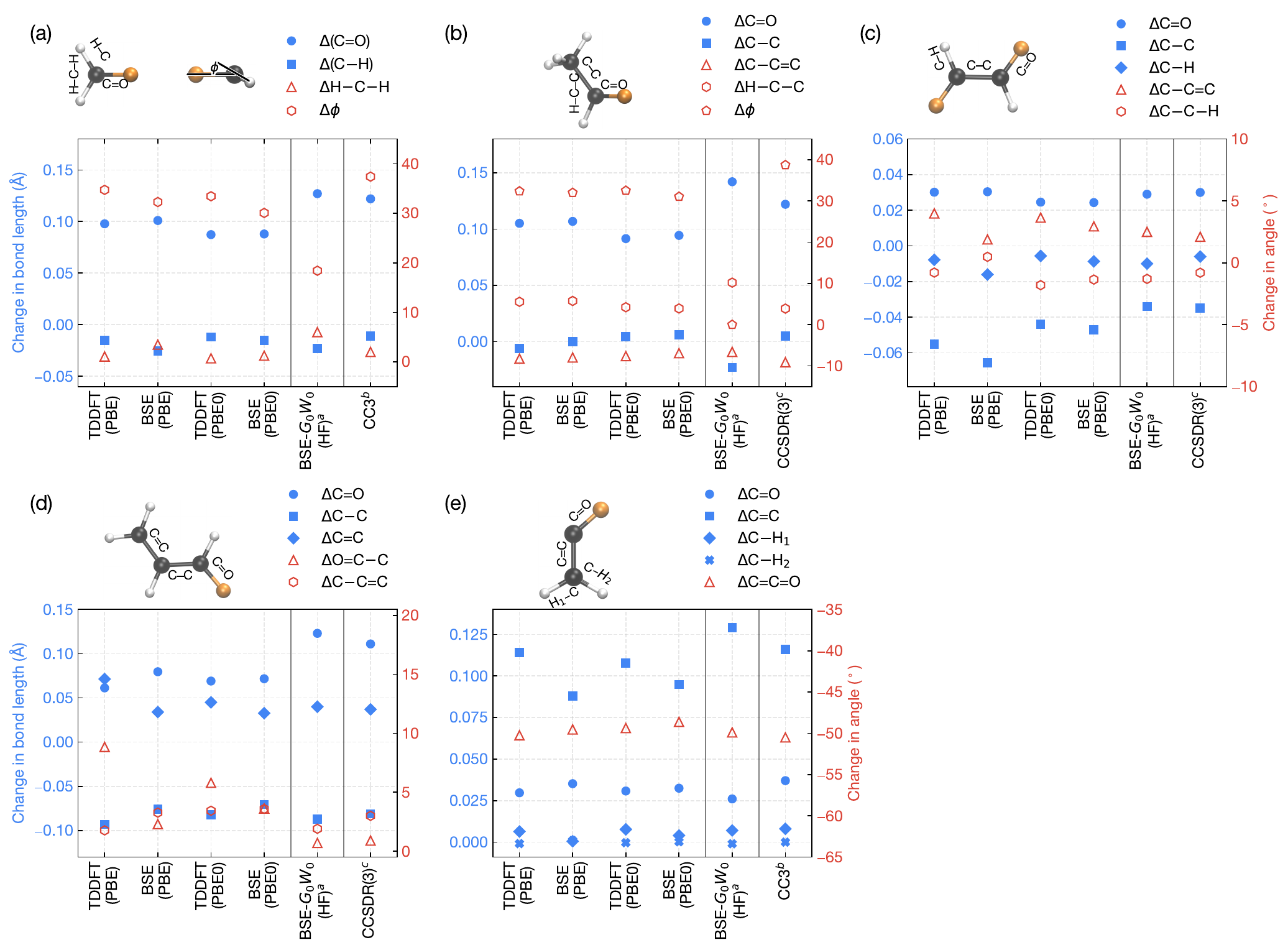}
    \caption{Change of bond lengths and angles in the optimized first singlet excited-state geometry relative to the ground-state geometry for (a) formaldehyde, (b) acetaldehyde, (c) glyoxal, (d) acrolein, and (e) ketene. \textsuperscript{a}BSE-$G_0W_0$(HF) geometries are from Ref.~\citenum{tolle2025fully}. \textsuperscript{b}CC3 and \textsuperscript{c}CCSDR(3) reference values are taken from Ref.~\citenum{budzak2017accurate}.}
    \label{s-fig:molecules}
\end{figure}

We further tested the analytical BSE forces by optimizing excited-state geometries for several small molecules, as shown in Fig.~\ref{s-fig:molecules}. We compare the changes in bond lengths and angles relative to the ground-state geometries obtained from TDDFT and BSE calculations with PBE~\cite{perdew1996generalized} and PBE0~\cite{adamo1999toward} starting points. These results are benchmarked against high-level quantum-chemistry reference values, as well as recent geometries obtained using fully analytical BSE-$G_0W_0$ forces with a Hartree-Fock (HF) starting point~\cite{tolle2025fully}. 

Overall, the TDDFT and BSE optimized excited-state geometries show similar trends for both PBE and PBE0 starting points, indicating that the present BSE forces implementation yields physically consistent structural relaxations across this molecular test set. The differences between TDDFT and BSE are generally smaller than those associated with the choice of starting functional, with PBE0 often giving geometrical changes closer to the high-level quantum-chemistry references. In some cases, BSE improves upon the corresponding TDDFT results, for example for the $\Delta$O$=$C--C angle in acrolein [Fig.~\ref{s-fig:molecules}(d)]. Compared with the fully analytical BSE-$G_0W_0$ results obtained from a Hartree-Fock starting point~\cite{tolle2025fully}, our BSE geometries from DFT starting points reproduce the same qualitative relaxation patterns. In several cases, such as the out-of-plane angle $\Delta \phi$ in formaldehyde and acetaldehyde [Fig.~\ref{s-fig:molecules}(a,b)], the DFT-starting-point BSE results are in much closer agreement with the CC3 and CCSDR(3) reference values~\cite{budzak2017accurate}. This comparison suggests that the mean-field starting point can have an important influence on BSE excited-state geometries. These benchmarks support the accuracy of our analytical BSE forces implementation and show that the approximations used here provide reliable excited-state structural relaxations for small molecules.

\newpage
\section{Computational parameters for the point-defect calculations}

DFT calculations were performed within the plane-wave pseudopotential framework implemented in the \textsc{Quantum ESPRESSO} package~\cite{giannozzi2020qe,carnimeo2023quantum}. The plane-wave kinetic-energy cutoff was set to 60 Ry, and optimized norm-conserving Vanderbilt (ONCV) pseudopotentials were used~\cite{hamann2013optimized,schlipf2015optimization}. We used the semilocal Perdew--Burke--Ernzerhof (PBE) functional~\cite{perdew1996generalized} and, for NV$^-$ in diamond, a dielectric-dependent hybrid (DDH) functional with an exact-exchange (EXX) fraction of 0.18~\cite{skone2014self,skone2016nonempirical}. For C$_{\mathrm{B}}$C$_{\mathrm{N}}^0$ in 2D-hBN, we used a PBE0-type hybrid functional with an EXX fraction of 0.41, obtained by enforcing the generalized Koopmans' condition for localized defect states~\cite{smart2018fundamental}; this functional is denoted as DDH$\alpha$ in this work. Unless otherwise specified, a $(4\times4\times4)$ 511-atom supercell was used for NV$^-$ in diamond, and a $(12\times12\times1)$ 288-atom supercell was used for C$_{\mathrm{B}}$C$_{\mathrm{N}}^0$ in 2D-hBN. A vacuum spacing of 20 \AA\ was used for the 2D-hBN supercell. The Brillouin zone of each supercell was sampled at the $\Gamma$ point. The ground state geometries were relaxed using DFT with both the PBE and DDH (DDH$\alpha$) functionals until the maximum absolute value of the ground-state forces component was smaller than 0.01 eV/\AA.

TDDFT and BSE calculations were performed with TDA using the \textsc{WEST} code~\cite{govoni2015large,yu2022gpu,jin2023excited,yu2024gpu}. Excited-state geometries were relaxed until the maximum absolute value of the excited-state forces component was smaller than 0.03 eV/\AA. The statically screened Coulomb interaction is evaluated using the projective dielectric eigenpotential (PDEP) technique within the random-phase approximation, with the number of the PDEP eigenpotentials set equal to two times the number of occupied orbitals. In BSE calculations using a scissor operator, the scissor value was determined by matching the vertical excitation energy at the ground-state geometry to that obtained from the corresponding BSE-$G_0W_0$ calculation. For NV$^-$ in diamond, the scissor value is 1.17 eV (0.45 eV) when using a PBE (DDH) functional starting point. For C$_{\mathrm{B}}$C$_{\mathrm{N}}^0$ in 2D-hBN, the scissor value is 2.77 eV (1.01 eV) when using a PBE (DDH$\alpha$) functional starting point.

Vibrational modes were computed in the electronic ground state using the frozen-phonon approach using the PBE functional for the save of computational cost. The displaced configurations were generated with the \textsc{Phonopy} package~\cite{phonopy-phono3py-JPCM,phonopy-phono3py-JPSJ}, using a displacement amplitude of 0.01 \AA\ from the PBE equilibrium geometry. To account for finite-size effects for NV$^-$ in diamond, the phonon modes were extrapolated to the dilute limit, approximated by a $(12\times12\times12)$ supercell containing 13,823 atoms, following the force-constant-matrix embedding approach proposed by Alkauskas \textit{et al.}~\cite{alkauskas2014first,razinkovas2021vibrational}.

The vibrationally resolved photoluminescence (PL) spectra were computed within Huang--Rhys theory~\cite{huang1950theory}, using the generating-function approach implemented in the \textsc{PyPL} package~\cite{alkauskas2014first,jin2021photoluminescence,jin2022vibrationally}. The PL line shape is written as
\begin{equation}
    I(\hbar\omega, T) \propto (\hbar\omega)^3 A_{\mathrm{emi}}(E_{\mathrm{ZPL}} - \hbar\omega, T),
\end{equation}
where $E_{\mathrm{ZPL}}$ is the zero-phonon-line energy, $\hbar\omega$ is the emitted photon energy, and $T$ is the temperature. The emission spectral function is
\begin{equation}
    A_{\mathrm{emi}}(\hbar\omega, T) = \frac{1}{2\pi} \int_{-\infty}^{\infty} e^{i\omega t} G_{\mathrm{emi}}(t,T) e^{-\frac{\lambda |t|}{\hbar}} dt,
\end{equation}
where $\lambda$ is a broadening parameter. The generating function is
\begin{equation}
\begin{aligned}
    G_{\mathrm{emi}}(t,T) &= \exp\bigg[ \int_{-\infty}^{\infty} S(\hbar\omega) e^{-i\omega t} d(\hbar\omega) - \sum_k S_k \\ 
    &\quad + \int_{-\infty}^{\infty} C(\hbar\omega,T) e^{-i\omega t} d(\hbar\omega) + \int_{-\infty}^{\infty} C(\hbar\omega,T) e^{i\omega t} d(\hbar\omega) - 2\sum_k \overline{n}_k(T)S_k \bigg].
\end{aligned}
\end{equation}
Here, $\overline{n}_k(T)$ is the average occupation number of the $k$-th phonon mode. The spectral densities of electron-phonon coupling are defined as
\begin{equation}
\begin{aligned}
    S(\hbar\omega) = \sum_k S_k \delta(\hbar\omega-\hbar\omega_k), \quad C(\hbar\omega,T) = \sum_k \overline{n}_k(T) S_k \delta(\hbar\omega-\hbar\omega_k).
\end{aligned}
\end{equation}
In practical calculations, the $\delta$ functions are replaced by Gaussian functions to account for the continuum of phonon modes participating in the optical transition. The Huang--Rhys factor for mode $k$ is
\begin{equation}
S_k = \frac{\omega_k \Delta Q_k^2}{2\hbar},
\end{equation}
where $\Delta Q_k$ is the mass-weighted displacement along the $k$-th mode. It can be evaluated from the atomic displacements between the excited- and ground-state geometries,
\begin{equation}
    \Delta Q_k = \sum_{\alpha=1}^{N_{\mathrm{atom}}} \sum_{i=x,y,z} \sqrt{M_\alpha} \left( R_{\alpha i}^{\mathrm{ES}} - R_{\alpha i}^{\mathrm{GS}} \right) e_{k,\alpha i},
\end{equation}
or, equivalently within the harmonic approximation, from the ground-state forces evaluated at the relaxed excited-state geometry~\cite{alkauskas2014first,razinkovas2021vibrational},
\begin{equation}
\Delta Q_k =
\frac{1}{\omega_k^2}
\sum_{\alpha=1}^{N_{\mathrm{atom}}}
\sum_{i=x,y,z}
\frac{F_{\alpha i}}{\sqrt{M_\alpha}}
e_{k,\alpha i}.
\end{equation}
Here, $M_\alpha$ is the mass of atom $\alpha$, and $\omega_k$ and $\mathbf{e}_k$ are the frequency and eigenvector of phonon mode $k$, respectively.

\section{Additional details on the carbon-dimer defect in 2D-hBN}

\begin{figure}[htp!]
\centering
\includegraphics[width=0.8\linewidth]{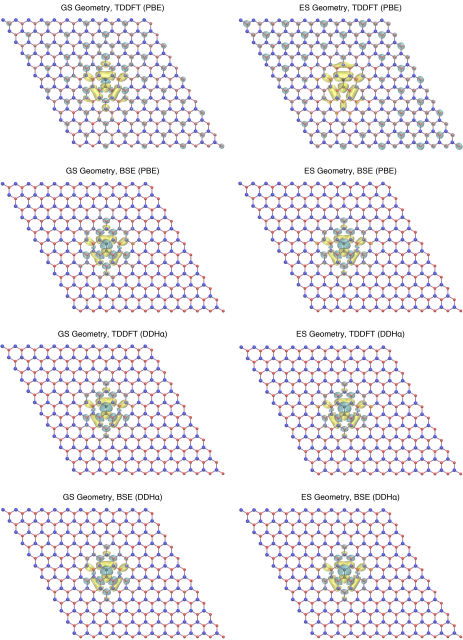}
\caption{Unrelaxed differential density computed using TDDFT-PBE (first row), BSE-PBE (second row), TDDFT-DDH$\alpha$ (third row), and BSE-DDH$\alpha$ (fourth row), evaluated at the ground-state geometry (first column) and at the relaxed excited-state geometry (second column). Yellow and blue indicate density accumulation and depletion, respectively.}
\label{s-fig:diff_rho_3d}
\end{figure}

To further understand the differences between TDDFT and BSE in describing the excited state and geometry relaxation of the C$_{\mathrm{B}}$C$_{\mathrm{N}}^0$ defect in 2D-hBN, we compute the unrelaxed differential density, defined in Eq.~(5) of the main text, at both the ground-state and relaxed excited-state geometries. The results are shown in Fig.~\ref{s-fig:diff_rho_3d}. The differential density describes the change in electron density upon vertical excitation at a fixed nuclear geometry. At the ground-state geometry, the differential densities obtained with different methods show broadly similar features, with density accumulation and depletion localized around the carbon dimer. However, the density depletion obtained from TDDFT-PBE is noticeably more delocalized than in the other approaches, suggesting that the excitation already contains mixed character from a transition involving delocalized valence-band states of 2D-hBN and the localized unoccupied defect orbital.

At the relaxed excited-state geometry obtained with each level of theory, the TDDFT-PBE differential density becomes qualitatively different from those obtained with BSE-PBE, TDDFT-DDH$\alpha$, and BSE-DDH$\alpha$. In particular, the density accumulation is localized around the defect center, whereas the density depletion is strongly delocalized across the supercell. This indicates that the relaxed TDDFT-PBE excited state is dominated by a transition from a delocalized occupied valence-band state of 2D-hBN to the localized unoccupied defect orbital. This finding is consistent with the substantially different mass-weighted displacement, C--C, C--B, and C--N bond-length changes, and PL line shape predicted by TDDFT-PBE shown in Fig.~3 of the main text. The qualitatively different relaxation can be attributed to the absence of an explicit screened electron-hole attraction in TDDFT with semilocal functionals, which is included through the $\mathcal{K}^{\mathrm{1d}}$ term in BSE. In the DDH$\alpha$ calculations, the inclusion of exact exchange introduces an analogous exchange-driven electron-hole interaction through the $\mathcal{K}^{\mathrm{1d,KS}}$ term, improving the localization of the defect excitation.

In contrast, the differential densities from BSE-PBE, TDDFT-DDH$\alpha$, and BSE-DDH$\alpha$ remain localized and visually similar at both the ground-state and relaxed excited-state geometries. This indicates that these three approaches describe the same qualitative defect excitation. The similarity is consistent with the corresponding excited-state relaxation patterns, as reflected in the changes of the C--C, C--B, and C--N bond lengths.

\begin{figure}[htp!]
\centering
\includegraphics[width=0.8\linewidth]{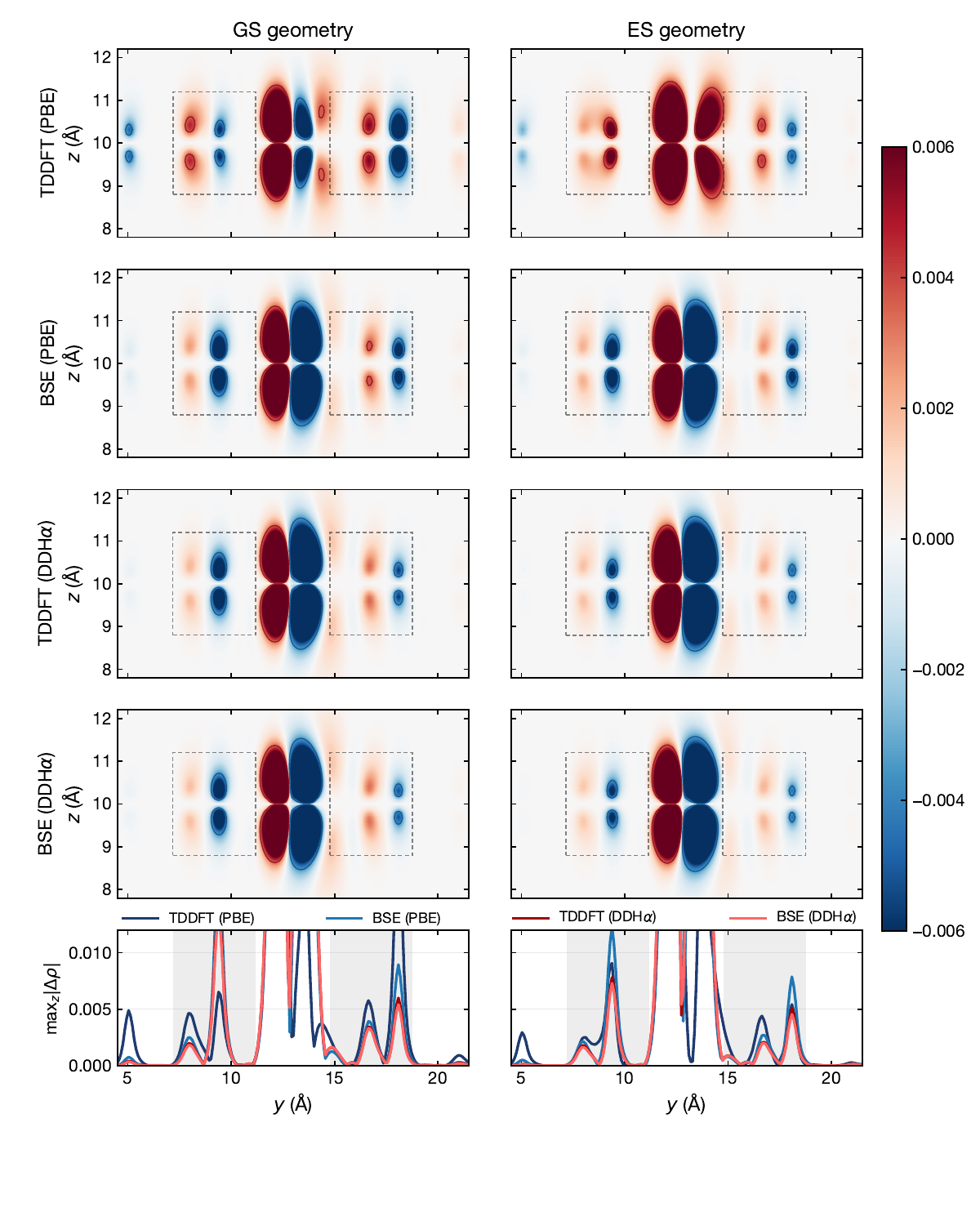}
\caption{Contour plots of the unrelaxed differential density in the $yz$ plane containing the C--C bond. The bottom panel shows the maximum absolute value of the differential density along the $z$ direction as a function of the $y$ position for different methods.}
\label{s-fig:diff_rho_2d}
\end{figure}

To examine the differences between BSE-PBE and BSE-DDH$\alpha$ more closely, we plot contour cuts of the differential density in the $yz$ plane containing the C--C bond in Fig.~\ref{s-fig:diff_rho_2d}. The TDDFT-PBE differential density is again qualitatively distinct from the other cases, especially at the relaxed excited-state geometry. The BSE-PBE differential density is slightly more delocalized than the BSE-DDH$\alpha$ result, as seen more clearly in the bottom panel of Fig.~\ref{s-fig:diff_rho_2d}, which shows the maximum absolute value of the differential density along the $z$ direction as a function of the $y$ position. This slightly more delocalized character is consistent with the smaller excited-state relaxation and weaker phonon sideband obtained with BSE-PBE. Because the Huang--Rhys factors depend quadratically on the mode-resolved displacements, even modest differences in the relaxed geometry can lead to more pronounced differences in the PL line shape. These differences reflect the remaining starting-point dependence of the BSE calculation.

The TDDFT-DDH$\alpha$ and BSE-DDH$\alpha$ differential densities are nearly identical, consistent with their very similar excited-state relaxations and PL line shapes.

\begin{table}[htp!]
    \centering
    \caption{GPU hours required for individual TDDFT and BSE forces evaluations with both the PBE and DDH functional starting point for NV$^-$ in diamond and C$_\mathrm{B}$C$_\mathrm{N}^{0}$ in 2D-hBN. All calculations were performed on NVIDIA A100 GPUs on NERSC Perlmutter.}
    \begin{tabular*}{0.6\columnwidth}{@{\extracolsep{\fill}}lcccc}
    \hline\hline \\ [-8pt]
    & \multicolumn{2}{c}{TDDFT} & \multicolumn{2}{c}{BSE} \\ [2pt]
    & PBE & DDH & PBE & DDH \\ [2pt]
    \hline \\ [-8pt]
    NV$^-$ in diamond & 2.4 & 55 & 175 & 500 \\ [2pt]
    \hline \\ [-8pt]
    C$_{\mathrm{B}}$C$_{\mathrm{N}}^0$ in 2D-hBN & 1.9 & 41 & 99 & 178 \\ [2pt]
    \hline\hline
    \end{tabular*}
    \label{s-tab:GPU hours}
\end{table}

\newpage
\section{Comparison of excitation energies with previous studies}

For completeness, we provide in Table~\ref{tab:vee_aee} a comparison of the vertical excitation energies (VEEs) and adiabatic excitation energies (AEEs) of NV$^-$ in diamond and C$_{\mathrm{B}}$C$_{\mathrm{N}}^0$ in 2D-hBN. The table includes the TDDFT and BSE results obtained in this work, together with previous theoretical and experimental values.

\begin{table}[ht]
\caption{Vertical excitation energies (VEEs) and adiabatic excitation energies (AEEs) of NV$^-$ in diamond and C$_{\mathrm{B}}$C$_{\mathrm{N}}^{0}$ in 2D hBN. For C$_{\mathrm{B}}$C$_{\mathrm{N}}^{0}$, DDH denotes DDH$\alpha$. All energies are in eV.}
\label{tab:vee_aee}
\centering
\begin{tabular*}{\columnwidth}{@{\extracolsep{\fill}}lcccccc}
\hline \hline
 & \multicolumn{4}{c}{This work} & \multicolumn{2}{c}{Literature} \\
 & \multicolumn{2}{c}{TDDFT} & \multicolumn{2}{c}{BSE} & Theo. & Expt. \\
 & PBE & DDH & PBE & DDH & & \\ [2pt]
\hline \\ [-8pt]
\multicolumn{7}{c}{NV$^-$ in diamond} \\ [2pt]
VEE & 2.090 & 2.395 & 2.174 & 2.428 & 2.235\textsuperscript{a}, 2.10\textsuperscript{b}, 2.40\textsuperscript{c}, 2.11\textsuperscript{d}, 2.24\textsuperscript{e}& \\
AEE & 1.895 & 2.100 & 1.982 & 2.172 & 1.955\textsuperscript{f}, 1.80\textsuperscript{d} & 1.945\textsuperscript{g} \\

\hline \\ [-8pt]
\multicolumn{7}{c}{C$_{\mathrm{B}}$C$_{\mathrm{N}}^0$ in 2D-hBN} \\ [2pt]
VEE & 4.000 & 5.015 & 4.169 & 4.719 & 4.64\textsuperscript{h}, 4.76\textsuperscript{i}, 4.54\textsuperscript{b}, 4.53\textsuperscript{j}, 4.23\textsuperscript{k}, 4.78\textsuperscript{l} & \\
AEE & 3.780 & 4.861 & 4.087 & 4.567 & 4.49\textsuperscript{h}, 4.31\textsuperscript{j} & 4.1\textsuperscript{g} \\
\hline \hline
\end{tabular*}

\vspace{1mm}
\begin{minipage}{\columnwidth}
\raggedright
\hspace{0mm}\textsuperscript{a} QDET (DDH)~\cite{chen2025advances}.
\hspace{0mm}\textsuperscript{b} ppRPA (B3LYP)~\cite{li2024accurate}.
\hspace{0mm}\textsuperscript{c} DMET-NEVPT2~\cite{haldar2023local}.
\hspace{0mm}\textsuperscript{d} CASSCF-NEVPT2~\cite{benedek2025accurate}.
\hspace{0mm}\textsuperscript{e} Emb-NEVPT2~\cite{martirez2026optical}.

\hspace{0mm}\textsuperscript{f} $\Delta$SCF (HSE06)~\cite{gali2009theory}.
\hspace{0mm}\textsuperscript{g} Experimentally measured ZPL energy for NV$^-$ in diamond~\cite{davies1976optical} and C$_{\mathrm{B}}$C$_{\mathrm{N}}^0$ in hBN~\cite{museur2008defect,du2015origin}.

\hspace{0mm}\textsuperscript{h} VEE obtained using BSE-ev$GW$ (PBE0) while the excited state geometry is relaxed using TDDFT (PBE0) forces~\cite{winter2021photoluminescent}.

\hspace{0mm}\textsuperscript{i} EOM-CCSD~\cite{lau2024optical}. 
\hspace{0mm}\textsuperscript{j} $\Delta$SCF (HSE with $\alpha_{\mathrm{EXX}}=0.40$)~\cite{mackoit2019carbon}.
\hspace{0mm}\textsuperscript{k} cRPA (HSE with $\alpha_{\mathrm{EXX}}=0.40$)~\cite{muechler2022quantum}.

\hspace{0mm}\textsuperscript{l} TDDFT (CAM-B3LYP)~\cite{korona2019exploring}.
\end{minipage}
\end{table}


\bibliography{SM}